\documentclass{article}

\usepackage[preprint]{neurips_2019}

\usepackage[utf8]{inputenc} 
\usepackage[T1]{fontenc}    
\usepackage{hyperref}       
\usepackage{url}            
\usepackage{booktabs}       
\usepackage{amsfonts,amsmath}       
\usepackage{nicefrac}       
\usepackage{microtype}      
\usepackage{graphicx}
\usepackage[capitalise]{cleveref}
\usepackage{todonotes}
\usepackage{pifont}
\newcommand{\cmark}{\ding{51}}%
\newcommand{\xmark}{\ding{55}}%

\usepackage{hyperref}       
\usepackage{xcolor}
\usepackage{sidecap}
\hypersetup{
    colorlinks,
    linkcolor={blue!50!black},
    citecolor={blue!50!black},
    urlcolor={blue!50!black}
}

\title{Machine Learning with Multi-Site Imaging Data:\\An Empirical Study on the Impact of Scanner Effects}

\author{Anonymous}

\author{
Ben Glocker$^1$,
~ Robert Robinson$^1$,
~Daniel C. Castro$^1$,
~Qi Dou$^1$,
~Ender Konukoglu$^2$ \\
\\
$^1$ Biomedical Image Analysis Group, Imperial College London, UK \\
$^2$ Computer Vision Laboratory, ETH Zurich, Zurich, Switzerland
}

\begin{document}

\maketitle
\vspace{-5mm}
\begin{abstract}
\vspace{-3mm}
This is an empirical study to investigate the impact of scanner effects when using machine learning on multi-site neuroimaging data. We utilize structural T1-weighted brain MRI obtained from two different studies, Cam-CAN and UK Biobank. For the purpose of our investigation, we construct a dataset consisting of brain scans from 592 age- and sex-matched individuals, 296 subjects from each original study. Our results demonstrate that even after careful pre-processing with state-of-the-art neuroimaging pipelines a classifier can easily distinguish between the origin of the data with very high accuracy. Our analysis on the example application of sex classification suggests that current approaches to harmonize data are unable to remove scanner-specific bias leading to overly optimistic performance estimates and poor generalization. We conclude that multi-site data harmonization remains an open challenge and particular care needs to be taken when using such data with advanced machine learning methods for predictive modelling.
\end{abstract}

\section{Motivation}\label{sec:intro}
Pooling data from different sites and previous studies is essential for analysis of large populations with sufficient statistical power \citep{smith2018statistical}. However, due to differences in image acquisition, demographics, disease characteristics and other factors, naive combination of datasets for subsequent large-scale population analysis can be problematic. Here, we conduct a simple, empirical study to illustrate and highlight this problem in the context of machine learning. We are not suggesting a solution, but rather re-iterate that multi-center data harmonization is an open research challenge. For some recent attemps to tackle this problem, see for example \citep{fortin2017harmonization,fortin2018harmonization}.

\section{Data}\label{sec:data}

We construct an age- and sex-matched dataset with T1-weighted brain MRI from $n=592$ individuals, where $296$ subjects ($146$ females) are taken each from the Cambridge Centre for Ageing and Neuroscience study (Cam-CAN)\footnote{\url{http://www.cam-can.org/}} \citep{shafto2014cambridge,taylor2017cambridge} and UK Biobank imaging study (UKBB)\footnote{\url{http://www.ukbiobank.ac.uk/}} \citep{sudlow2015,miller2016,alfaro-almagro2018}. This is to simulate a somewhat `best case scenario' for multi-site data where the age- and sex-matching intends to remove population bias. We note this is rarely possible in practice, and it is expected that current and previous analyses that pool data from different sites suffer from much larger site-specific biases.

\paragraph{Cam-CAN:} All images were collected at a single site (Medical Research Council Cognition and Brain Sciences Unit (MRC-CBSU) in Cambridge, UK) using a 3T Siemens TIM Trio scanner with a 32-channel receive head coil. Imaging parameters are: 3D MPRAGE, TR=2250ms, TE=2.99ms, TI=900ms; FA=9 deg; FOV=256x240x192mm; 1mm isotropic; GRAPPA=2; TA=4mins 32s.

\paragraph{UK Biobank:} All images were collected at the UKBB imaging center using a 3T Siemens Skyra scanner with a 32-channel receive head coil. Imaging parameters are: 3D MPRAGE, R=2, TR=2000ms, TE=385ms, TI=880ms; FOV=208x256x256mm; 1mm isotropic; Duration 4mins 54s.

The acquisition protocols of the two studies are remarkably similar, and possibly much closer than typically found when pooling data from multiple sites. The subjects in both studies should be normal.

\subsection{Pre-Processing Pipeline}\label{sec:preproc}

We aimed at designing a common state-of-the-art pre-processing pipepline which in this or similar form is widely used in neuroimaging studies. In particular, we apply the following sequential steps: 1) Lossless image reorientation by swapping axes using the direction information from the NIfTI image header, such that all scans are in the same radiological orientation of left, posterior, superior; 2) Skull stripping with ROBEX v1.2\footnote{\url{https://www.nitrc.org/projects/robex}} \citep{iglesias2011robust}; 3) Bias field correction with N4ITK\footnote{\url{https://itk.org}} \citep{tustison2010n4itk}; 4) Intensity-based linear registration (rigid and affine) to MNI ICBM 152 2009a Nonlinear Symmetric\footnote{\url{http://nist.mni.mcgill.ca/?p=904}} using an in-house registration tool with correlation coefficient as the similarity measure and downhill-simplex as the optimizer.

After these steps, we perform intensity normalization within brain regions with simple whitening (zero-mean/unit-variance). Voxels outside the brain are set to fixed value. Other techniques such as percentile matching and Nyul's histogram standardization \citep{nyul2000new} led to similar subsequent observations. We also employ SPM12\footnote{\url{http://www.fil.ion.ucl.ac.uk/spm/software/spm12/}} \citep{spm_book,ashburner2012spm} and FMRIB's Automated Segmentation Tool (FAST) v4.0\footnote{\url{https://fsl.fmrib.ox.ac.uk/fsl/fslwiki/FAST}} \citep{zhang2001segmentation} to obtain brain tissue probability maps. SPM is run directly on the raw T1-weighted scans as it has its own pre-processing pipeline built-in including spation non-linear normalization to MNI space. FSL-FAST is run on our skull-stripped, bias field corrected and rigidly MNI aligned images.


\begin{SCfigure}
	\centering
    \includegraphics[width=.6\linewidth]{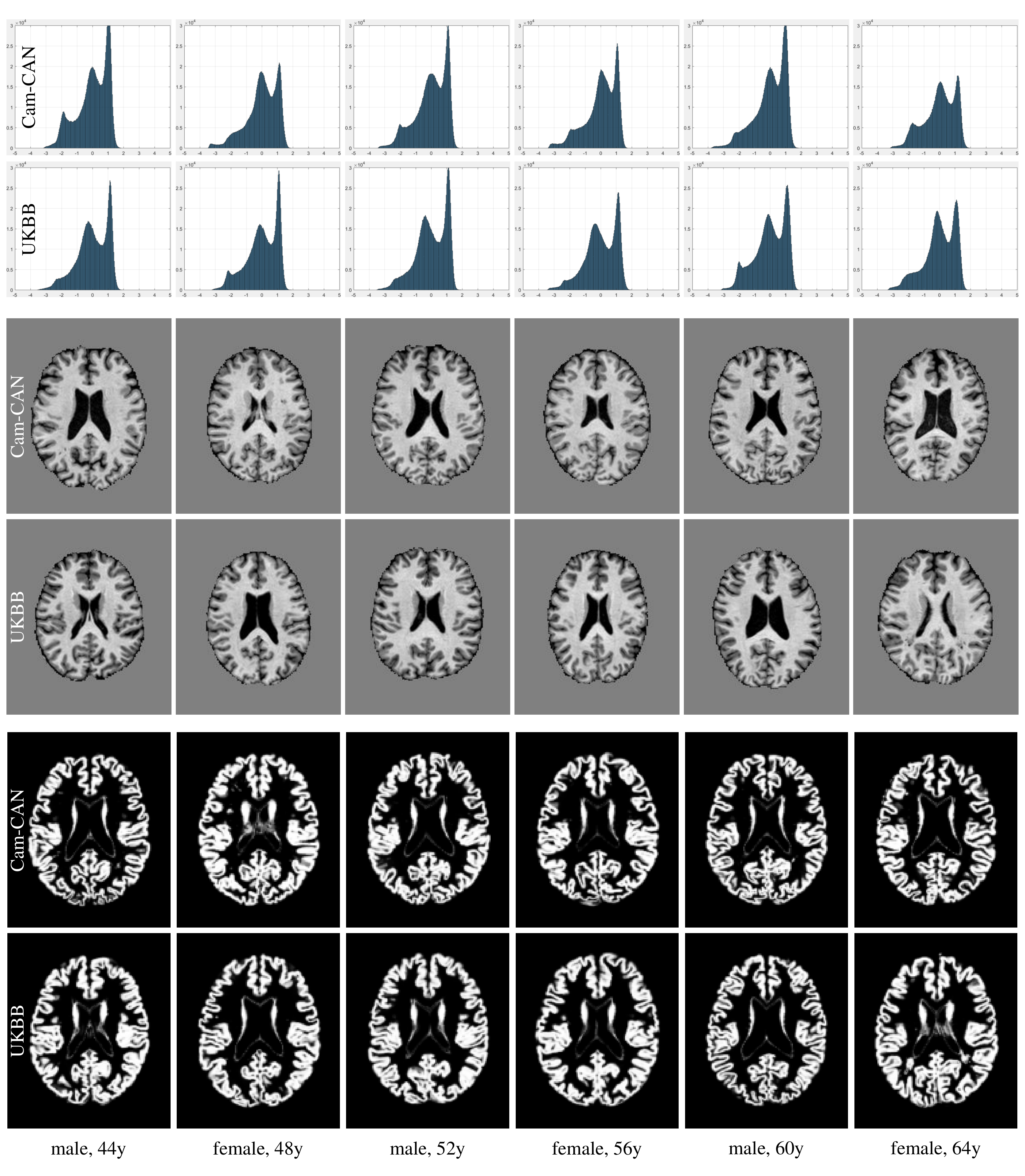}
    \caption{Example data for six age- and sex-matched subjects from the Cam-CAN and UKBB datasets after applying different pre-processing steps. Top two rows show the intensity histograms after skull-stripping, bias field correction, rigid registration to MNI, and whitening for intensity normalization. Rows three and four show the corresponding T1-weighted mid axial slices. Rows five and six show the spatially normalized graymatter maps obtained with SPM12. Site-specific differences are non-obvious from visual inspection.}
    \label{fig:data_overview}
\end{SCfigure}

\section{Experiments, Results \& Conclusion}\label{sec:experiments}

We conduct two image classification experiments to illustrate the impact of scanner effects which remain after careful pre-processing and are present even in image-derived tissue probability maps.

\textbf{Site classification:} We train random forest binary classifiers to distinguish between the origin of the imaging data. The classifiers are trained to distinguish between data from Cam-CAN and UKBB.

Results are summarized in Table~\ref{tab:site-classification}. We make the following observations: i) classifiers are able to predict data origin with high accuracy; ii) scanner effects remain in derived tissue probability maps; iii) higher degrees of spatial normalization amplify scanner effects (possibly related to interpolation).

\textbf{Sex classification:} We consider a simple binary classification task of sex classification. We compare results of training random forest classifiers on single-site and multi-site data.

Results for sex classification are summarized in Table~\ref{tab:sex-classification}. We make the following observations: i) age/sex-matched multi-site data gives realistic estimates of accuracy (similar to single site); ii) sex imbalance in multi-site leads to overly optimistic accuracy; iii) training on one site and testing on the other shows drop of performance indicating poor generalization; iv) when discriminative features such as brain size are removed by affine registration, the drop in performance is more severe.

\textbf{Conclusions:} Scanner effects can be subtle yet significantly affect machine learning. Similar findings for multi-site neuroimaging data are reported in \citep{ferrari2018pitfalls, wachinger2019quantifying}.

\begin{table}
  \begin{center}
    \caption{Two-fold cross validation results for site classification. Reported are overall accuracy, average entropy, and average predictive probability. If the data were indistinguishable one would expect an accuracy of 50$\%$, an entropy of 0.6931 (upper bound), and a probability of 0.5.}
    \label{tab:site-classification}
    \begin{tabular}{ccccccc}
      \toprule
      \textbf{Stripped} & \textbf{Bias Field} & \textbf{Aligned} & \textbf{Intensities} & \textbf{Accuracy} & \textbf{Avg. Entropy} & \textbf{Avg. Prob.}  \\
      \midrule
      \cmark & \cmark & rigid & whitening & 96.96\% & 0.4039 & 0.8296 \\
      \cmark & \cmark & affine & whitening & 98.82\% & 0.3876 & 0.8397 \\
      \midrule
      \multicolumn{4}{l}{\textbf{SPM12 -- Segment}} & \textbf{Accuracy} & \textbf{Avg. Entropy} & \textbf{Avg. Prob.}  \\
      \midrule      
      \xmark & \cmark & rigid & graymatter & 80.24\% & 0.6363 & 0.6399 \\
      \xmark & \cmark & non-linear & graymatter & 96.62\% & 0.5675 & 0.7234 \\
      \midrule
      \multicolumn{4}{l}{\textbf{FSL -- FAST}} & \textbf{Accuracy} & \textbf{Avg. Entropy} & \textbf{Avg. Prob.}  \\
      \midrule      
      \cmark & \cmark & rigid & graymatter & 93.24\% & 0.4542 & 0.7968 \\
      \bottomrule
    \end{tabular}
  \end{center}
\end{table}

\begin{table}
  \begin{center}
    \caption{Two-fold cross validation results for sex classification under different data arrangements.}
    \label{tab:sex-classification}
    \begin{tabular}{ccccccc}
      \toprule
      \multicolumn{3}{l}{\textbf{Data Arrangement}} & \textbf{Aligned} & \textbf{Accuracy} & \textbf{Avg. Entropy} & \textbf{Avg. Prob.}  \\
      \midrule
      \multicolumn{3}{l}{Multi-site age/sex-matched} & rigid & 82.60\% & 0.5304 & 0.7388 \\
      \multicolumn{3}{l}{Single-site (Cam-CAN)} & rigid & 81.42\% & 0.5592 & 0.7179 \\
      \multicolumn{3}{l}{Single-site (UKBB)} & rigid & 84.46\% & 0.5049 & 0.7572 \\
      \midrule
      \multicolumn{3}{l}{Cam-CAN females / UKBB males} & rigid & 94.59\% & 0.4036 & 0.8311 \\
      \multicolumn{3}{l}{Cam-CAN 80/20\% / UKBB 20/80\%} & rigid & 85.87\% & 0.5038 & 0.7616 \\
      \midrule
      \multicolumn{3}{l}{Cam-CAN train / UKBB test} & rigid & 81.42\% & 0.5617 & 0.7124 \\
      \multicolumn{3}{l}{UKBB train / Cam-CAN test} & rigid & 78.04\% & 0.5284 & 0.7419 \\
      \midrule
      \multicolumn{3}{l}{Multi-site age/sex-matched} & affine & 79.73\% & 0.6345 & 0.6389 \\
      \multicolumn{3}{l}{Single-site (Cam-CAN)} & affine & 77.70\% & 0.6439 & 0.6269 \\
      \multicolumn{3}{l}{Single-site (UKBB)} & affine & 81.08\% & 0.6393 & 0.6316 \\
      \midrule
      \multicolumn{3}{l}{Cam-CAN females / UKBB males} & affine & 98.99\% & 0.4641 & 0.8013 \\
      \multicolumn{3}{l}{Cam-CAN 80/20\% / UKBB 20/80\%} & affine & 84.78\% & 0.5713 & 0.7125 \\
      \midrule
      \multicolumn{3}{l}{Cam-CAN train / UKBB test} & affine & 73.65\% & 0.6462 & 0.6245 \\
      \multicolumn{3}{l}{UKBB train / Cam-CAN test} & affine & 62.16\% & 0.6075 & 0.6769 \\
      \bottomrule
    \end{tabular}
  \end{center}
\end{table}

\section*{Acknowledgements}
This research has received funding from the European Research Council (ERC) under the European Union's Horizon 2020 research and innovation programme (grant agreement No 757173, project MIRA, ERC-2017-STG). UK Biobank data has been accessed under Application Number 12579.

\bibliographystyle{hapalike}
\bibliography{references}

\end{document}